\documentclass [11pt]{article}
\usepackage[T1]{fontenc}
\usepackage[final]{epsfig}
\begin{document}

%
%
%
\edef\catcodeat{\the\catcode`\@ }     \catcode`\@=11
\newbox\p@ctbox                       
\newbox\t@mpbox                       
\newbox\@uxbox                        
\newbox\s@vebox                       
\newtoks\desct@ks \desct@ks={}        
\newtoks\@ppenddesc                   
\newtoks\sh@petoks                    
\newif\ifallfr@med  \allfr@medfalse   
\newif\if@ddedrows                    
\newif\iffirstp@ss  \firstp@ssfalse   
\newif\if@mbeeded                     
\newif\ifpr@cisebox                   
\newif\ifvt@p                         
\newif\ifvb@t                         
\newif\iff@nished    \f@nishedtrue    
\newif\iffr@med                       
\newif\ifj@stbox     \j@stboxfalse    
\newcount\helpc@unt                   
\newcount\p@ctpos                     
\newdimen\r@leth      \r@leth=0.4pt   
\newdimen\x@nit                       
\newdimen\y@nit                       
\newdimen\xsh@ft                      
\newdimen\ysh@ft                      
\newdimen\@uxdimen                    
\newdimen\t@mpdimen                   
\newdimen\t@mpdimeni                  
\newdimen\b@tweentandp                
\newdimen\b@ttomedge                  
\newdimen\@pperedge                   
\newdimen\@therside                   
\newdimen\d@scmargin                  
\newdimen\p@ctht                      
\newdimen\l@stdepth
\newdimen\in@tdimen
\newdimen\l@nelength
\newdimen\re@lpictwidth
\def\justframes{\global\j@stboxtrue}  
\def\picturemargins#1#2{\b@tweentandp=#1\@therside=#2\relax}
\def\allframed{\global\allfr@medtrue} 
\def\emptyplace#1#2{\pl@cedefs        
    \setbox\@uxbox=\vbox to#2{\n@llpar
        \hsize=#1\vfil \vrule height0pt width\hsize}
    \e@tmarks}
\def\boxplace{\pl@cedefs\afterassignment\re@dvbox\let\n@xt= }
\def\re@dvbox{\setbox\@uxbox=\vbox\bgroup
         \n@llpar\aftergroup\e@tmarks}
\def\fontcharplace#1#2{\pl@cedefs     
    \setbox\@uxbox=\hbox{#1\char#2\/}%
    \xsh@ft=-\wd\@uxbox               
    \setbox\@uxbox=\hbox{#1\char#2}%
    \advance\xsh@ft by \wd\@uxbox     
    \helpc@unt=#2
    \advance\helpc@unt by -63         
    \x@nit=\fontdimen\helpc@unt#1%
    \advance\helpc@unt by  20         
    \y@nit=\fontdimen\helpc@unt#1%
    \advance\helpc@unt by  20         
    \ifnum\helpc@unt<51
      \ysh@ft=-\fontdimen\helpc@unt#1%
    \fi
    \e@tmarks}
\def\n@llpar{\parskip0pt \parindent0pt
    \leftskip=0pt \rightskip=0pt
    \everypar={}}
\def\pl@cedefs{\xsh@ft=0pt\ysh@ft=0pt}
\def\e@tmarks#1{\setbox\@uxbox=\vbox{ 
      \n@llpar
      \hsize=\wd\@uxbox               
      \noindent\copy\@uxbox           
      \kern-\wd\@uxbox                
      #1\par}
    \st@redescription}
\def\t@stprevpict#1{\ifvoid#1\else    
   \errmessage{Previous picture is not finished yet.}\fi} 

\def\st@redescription#1\par{
    \global\setbox\s@vebox=\vbox{\box\@uxbox\unvbox\s@vebox}%
    \desct@ks=\expandafter{\the\desct@ks#1\@ndtoks}}
\def\def@ultdefs{\p@ctpos=1         
      \def\lines@bove{0}
      \@ddedrowsfalse               
      \@mbeededfalse                
      \pr@ciseboxfalse
      \vt@pfalse                    
      \vb@tfalse                    
      \@ppenddesc={}
      \ifallfr@med\fr@medtrue\else\fr@medfalse\fi
      }

\def\descriptionmargins#1{\global\d@scmargin=#1\relax}
\def\@dddimen#1#2{\t@mpdimen=#1\advance\t@mpdimen by#2#1=\t@mpdimen}
\def\placemark#1#2 #3 #4 #5 {\unskip    
      \setbox1=\hbox{\kern\d@scmargin#5\kern\d@scmargin}
      \@dddimen{\ht1}\d@scmargin        
      \@dddimen{\dp1}\d@scmargin        
      \ifx#1l\dimen3=0pt\else           
        \ifx#1c\dimen3=-0.5\wd1\else
          \ifx#1r\dimen3=-\wd1
     \fi\fi\fi
     \ifx#2u\dimen4=-\ht1\else          
       \ifx#2c\dimen4=-0.5\ht1\advance\dimen4 by 0.5\dp1\else
         \ifx#2b\dimen4=0pt\else
           \ifx#2l\dimen4=\dp1
     \fi\fi\fi\fi
     \advance\dimen3 by #3
     \advance\dimen4 by #4
     \advance\dimen4 by-\dp1
     \advance\dimen3 by \xsh@ft         
     \advance\dimen4 by \ysh@ft         
     \kern\dimen3\vbox to 0pt{\vss\copy1\kern\dimen4}
     \kern-\wd1                        
     \kern-\dimen3                     
     \ignorespaces}                    
\def\fontmark #1#2 #3 #4 #5 {\placemark #1#2 #3\x@nit{} #4\y@nit{} {#5} }
\def\fr@msavetopict{\global\setbox\s@vebox=\vbox{\unvbox\s@vebox
      \global\setbox\p@ctbox=\lastbox}%
    \expandafter\firstt@ks\the\desct@ks\st@ptoks}
\def\firstt@ks#1\@ndtoks#2\st@ptoks{%
    \global\desct@ks={#2}%
    \def\t@mpdef{#1}%
    \@ppenddesc=\expandafter\expandafter\expandafter
                        {\expandafter\t@mpdef\the\@ppenddesc}}
\def\testf@nished{{\let\s@tparshape=\relax
    \s@thangindent}}
\def\inspicture{\t@stprevpict\p@ctbox
    \def@ultdefs                  
    \fr@msavetopict
    \iff@nished\else\testf@nished\fi
    \iff@nished\else
      \immediate\write16{Previes picture is not finished yet}%
    \fi
    \futurelet\N@xt\t@stoptions}  
\def\t@stoptions{\let\n@xt\@neletter
  \ifx\N@xt l\p@ctpos=0\else                
   \ifx\N@xt c\p@ctpos=1\else               
    \ifx\N@xt r\p@ctpos=2\else              
     \ifx\N@xt(\let\n@xt\e@tline\else        
      \ifx\N@xt!\@mbeededtrue\else           
       \ifx\N@xt|\fr@medtrue\else            
        \ifx\N@xt^\vt@ptrue\vb@tfalse\else  
         \ifx\N@xt_\vb@ttrue\vt@pfalse\else 
          \ifx\N@xt\bgroup\let\n@xt\@ddgrouptodesc\else
           \let\n@xt\@dddescription 
  \fi\fi\fi\fi\fi\fi\fi\fi\fi\n@xt}
\def\e@tline(#1){\def\lines@bove{#1}
    \@ddedrowstrue
    \futurelet\N@xt\t@stoptions}
\def\@neletter#1{\futurelet\N@xt\t@stoptions} 
\def\@ddgrouptodesc#1{\@ppenddesc={#1}\futurelet\N@xt\t@stoptions}
\def\fr@medpict{\setbox\p@ctbox=
    \vbox{\n@llpar\hsize=\wd\p@ctbox
       \iffr@med\else\r@leth=0pt\fi
       \ifj@stbox\r@leth=0.4pt\fi
       \hrule height\r@leth \kern-\r@leth
       \vrule height\ht\p@ctbox depth\dp\p@ctbox width\r@leth \kern-\r@leth
       \ifj@stbox\hfill\else\copy\p@ctbox\fi
       \kern-\r@leth\vrule width\r@leth\par
       \kern-\r@leth \hrule height\r@leth}}
\def\@dddescription{\fr@medpict     
    \re@lpictwidth=\the\wd\p@ctbox
    \advance\re@lpictwidth by\@therside
    \advance\re@lpictwidth by\b@tweentandp
    \ifhmode\ifinner\pr@ciseboxtrue\fi\fi
    \createp@ctbox
    \let\N@xt\tr@toplacepicture
    \ifhmode                         
      \ifinner\let\N@xt\justc@py
      \else\let\N@xt\vjustc@py
      \fi
    \else
      \ifnum\p@ctpos=1               
        \let\N@xt\justc@py
      \fi
    \fi
    \if@mbeeded\let\N@xt\justc@py\fi 
    \firstp@sstrue
    \N@xt}
\def\createp@ctbox{\global\p@ctht=\ht\p@ctbox
    \advance\p@ctht by\dp\p@ctbox
    \advance\p@ctht by 6pt
    \setbox\p@ctbox=\vbox{
      \n@llpar                     
      \t@mpdimen=\@therside          
      \t@mpdimeni=\hsize             
      \advance\t@mpdimeni by -\@therside
      \advance\t@mpdimeni by -\wd\p@ctbox
      \ifpr@cisebox
        \hsize=\wd\p@ctbox
      \else
        \ifcase\p@ctpos
               \leftskip=\t@mpdimen    \rightskip=\t@mpdimeni
        \or    \advance\t@mpdimeni by \@therside
               \leftskip=0.5\t@mpdimeni \rightskip=\leftskip
        \or    \leftskip=\t@mpdimeni   \rightskip=\t@mpdimen
        \fi
      \fi
      \hrule height0pt             
      \kern6pt                     
      \penalty10000
      \noindent\copy\p@ctbox\par     
      \kern3pt                       
      \hrule height0pt
      \hbox{}%
      \penalty10000
      \interlinepenalty=10000
      \the\@ppenddesc\par            
      \penalty10000                  
      \kern3pt                       
      }%
      \ifvt@p
       \setbox\p@ctbox=\vtop{\unvbox\p@ctbox}%
      \else
        \ifvb@t\else
          \@uxdimen=\ht\p@ctbox
          \advance\@uxdimen by -\p@ctht
          {\vfuzz=\maxdimen
           \global\setbox\p@ctbox=\vbox to\p@ctht{\unvbox\p@ctbox}%
          }%
          \dp\p@ctbox=\@uxdimen
        \fi
      \fi
      }
\def\picname#1{\unskip\setbox\@uxbox=\hbox{\bf\ignorespaces#1\unskip\ }%
      \hangindent\wd\@uxbox\hangafter1\noindent\box\@uxbox\ignorespaces}
\def\justc@py{\ifinner\box\p@ctbox\else\kern\parskip\unvbox\p@ctbox\fi
  \global\setbox\p@ctbox=\box\voidb@x}
\def\vjustc@py{\vadjust{\kern0.5\baselineskip\unvbox\p@ctbox}%
      \global\setbox\p@ctbox=\box\voidb@x}
\def\tr@toplacepicture{
      \ifvmode\l@stdepth=\prevdepth  
      \else   \l@stdepth=0pt         
      \fi
      \vrule height.85em width0pt\par
      \r@memberdims                  
      \global\t@mpdimen=\pagetotal
      \t@stheightofpage              
      \ifdim\b@ttomedge<\pagegoal    
         \let\N@xt\f@gurehere        
         \global\everypar{}
      \else
         \let\N@xt\relax             
         \penalty10000
         \vskip-\baselineskip        
         \vskip-\parskip             
         \immediate\write16{Picture will be shifted down.}%
         \global\everypar{\sw@tchingpass}
      \fi
      \penalty10000
      \N@xt}
\def\sw@tchingpass{
    \iffirstp@ss                     
      \let\n@xt\relax
      \firstp@ssfalse                
    \else
      \let\n@xt\tr@toplacepicture
      \firstp@sstrue
    \fi  \n@xt}
\def\r@memberdims{\global\in@tdimen=0pt
    \ifnum\p@ctpos=0
        \global\in@tdimen=\re@lpictwidth
      \fi
      \global\l@nelength=\hsize
      \global\advance\l@nelength by-\re@lpictwidth
      }
\def\t@stheightofpage{%
     \global\@pperedge=\t@mpdimen
     \advance\t@mpdimen by-0.7\baselineskip 
     \advance\t@mpdimen by \lines@bove\baselineskip 
     \advance\t@mpdimen by \ht\p@ctbox      
     \advance\t@mpdimen by \dp\p@ctbox      
     \advance\t@mpdimen by-0.3\baselineskip 
     \global\b@ttomedge=\t@mpdimen          
     }
\def\f@gurehere{\global\f@nishedfalse
      \t@mpdimen=\lines@bove\baselineskip   
      \advance\t@mpdimen-0.7\baselineskip   
      \kern\t@mpdimen
      \advance\t@mpdimen by\ht\p@ctbox
      \advance\t@mpdimen by\dp\p@ctbox
      {\t@mpdimeni=\baselineskip
       \offinterlineskip
       \unvbox\p@ctbox
       \global\setbox\p@ctbox=\box\voidb@x
       \penalty10000   \kern-\t@mpdimen     
       \penalty10000   \vskip-\parskip      
       \kern-\t@mpdimeni                    
      }%
      \penalty10000                         
      \global\everypar{\s@thangindent}
      }
\def\s@thangindent{%
    \ifdim\pagetotal>\b@ttomedge\global\everypar{}%
      \global\f@nishedtrue             
      \else
        \advance\@pperedge by -1.2\baselineskip
        \ifdim\@pperedge>\pagetotal\global\everypar{}%
          \global\f@nishedtrue
        \else
          \s@tparshape                 
        \fi
        \advance\@pperedge by 1.2\baselineskip
      \fi}
\def\s@tparshape{\t@mpdimen=-\pagetotal
   \advance\t@mpdimen by\b@ttomedge    
   \divide\t@mpdimen by\baselineskip   
   \helpc@unt=\t@mpdimen               
   \advance \helpc@unt by 2            
   \sh@petoks=\expandafter{\the\helpc@unt\space}
   \t@mpdimeni=\lines@bove\baselineskip
   \t@mpdimen=\pagetotal
   \gdef\lines@bove{0}
   \loop \ifdim\t@mpdimeni>0.999\baselineskip 
     \advance\t@mpdimen  by \baselineskip
     \advance\t@mpdimeni by-\baselineskip
     \sh@petoks=\expandafter{\the\sh@petoks 0pt \the\hsize}%
   \repeat
   \loop \ifdim\b@ttomedge>\t@mpdimen         
     \advance\t@mpdimen by \baselineskip
     \sh@petoks=\expandafter{\the\sh@petoks \in@tdimen \l@nelength }%
   \repeat
   \sh@petoks=\expandafter
      {\the\sh@petoks 0pt \the\hsize}
   \expandafter\parshape\the\sh@petoks
   }

\descriptionmargins{2pt}
\picturemargins{15pt}{0pt}

\catcode`\@=\catcodeat        \let\catcodeat=\undefined

\emptyplace{3.3in \includegraphics{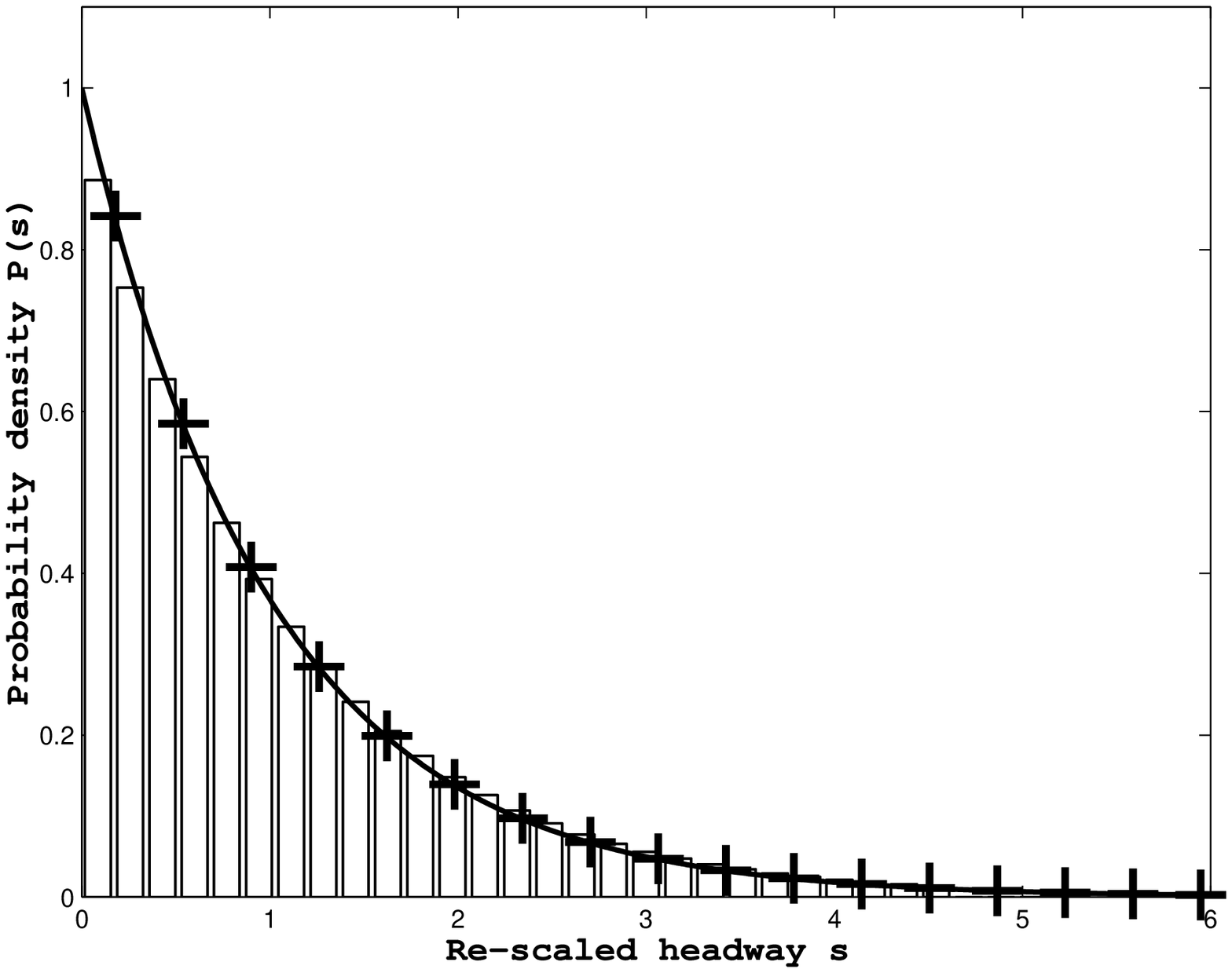}} {3.1in} {\footnotesize
\noindent FIGURE 1. Headway distribution of the ASEP . The
curve is (\ref{Poisson}). Bars and plus signs show the result
(\ref{Finalk}) after the re-scaling to mean headway equal to one
for $\alpha=0.15,\beta=0.6$ and $\alpha=0.3,\beta=0.5,$
respectively.}

\title{Headway distribution of the asymmetric simple exclusion model}
\author{Milan Krbalek$^{a,b}$
\\
    $^{a}$ Institute of Physics, Czech Academy of Science,\\
    Cukrovarnick\'a 10, Prague, Czech Republic\\
    $^{b}$ Faculty of Nuclear Sciences and Physical Engineering,\\
      Trojanova 13, Prague, Czech Republic\\
    e-mail: milan.krbalek@uhk.cz\\}

\maketitle
\begin{abstract}
We present an exact solution of headway distribution of the
asymmetric simple exclusion model with open boundary conditions
and compare it to the headway distributions of the highway
traffic.
\end{abstract}

\section{Introduction}
Recently, a new tendency has appeared in the physics of traffic
systems. Besides macroscopic characteristics of traffic (flow,
density, etc.), researchers are focused on microscopic properties
as well (\cite{Book}). One of these properties is the distribution
of the distances (or time intervals) between neighboring cars, so
called headway distribution. This function was investigated for
existing traffic models and compared to the various regimes of the
real traffic. It was observed that the real traffic data as well
as many traffic models lead to a headway distribution which
resembles the spacing distribution of the random-matrices
eigenvalues (\cite{Krbalek1},\cite{Krbalek2},\cite{Seba}). To
verify this observation, we investigate the asymmetric simple
exclusion process (ASEP) which is the well known exactly solvable
model of traffic (\cite{Derrida1},\cite{Derrida2},\cite{Santen}).
We find the exact formula for headway distribution of the
asymmetric simple exclusion model and compare it with the
eigenvalues distributions obtained from random matrix theory.


In the traffic theory, the spacing distribution which was often
called \textit{headway distribution} $P(s)$ is defined as
probability distribution of the distances $s$ between neighboring
cars (particles of the model). For the sake of simplicity we
assume that $\langle s \rangle \equiv \int_0^\infty s\,P(s)\,ds
=1.$ Among many others, we discern two special traffic systems in
real roads. The first is the \textit{bus transport} where no fixed
schedule exists (see \cite{Krbalek1},\cite{Seba}) and where the
interaction between bus drivers exists owing to the check points
in which the information about the time interval to the previous
bus is transmitted to a driver. It was found that the headway
distribution of that bus transport behaves as
\begin{equation}
P(s)=A\,s^2 \, e^{-B s^2}, \label{RMT-LS}
\end{equation}
where  $A,B$ are constants determined by conditions $\int_0^\infty
P(s) \, ds=1$ and $\int_0^\infty s\,P(s) \, ds=1.$ The equation
(\ref{RMT-LS}) originates from random matrix theory where it
describes the distribution of distances between eigenvalues of the
unitary random matrices.\\ The second system described here is
\textit{highway traffic} for which we distinguish three basic
regimes (\cite{Krbalek1}):\begin{itemize}
\item a) \textit{free-flow regime}, where the density of cars (number of
cars per unit of length) is low and the cars are therefore
moving as free. Headway of this regime is reduced to the
Poissonian distribution

\begin{equation}
P_0(s)=e^{-s}
\label{Poisson}
\end{equation}

\item b) \textit{synchronized regime}, where the mean velocity is
considerably reduced compared to the free-flow states due to a
higher density of cars, but cars are still moving (no traffic
jam). Headway distribution of these states is well described by
the formula

\begin{equation}
P_\nu(s)  =  \frac{(\nu+1)^{\nu+1}}{\Gamma(\nu +1)}s^{\nu}
e^{-(\nu +1) \, s}, \label{Bogo}
\end{equation}

where $\nu$ is a free parameter and $\Gamma$ is Gamma function.

\item c) \textit{stop-and-go regime}, where cars are moving in traffic
jams
\end{itemize}

Note that it is possible to obtain the formula (\ref{Poisson})
taking the limit $\nu \rightarrow 0^+$ in formula (\ref{Bogo}).
Distribution $P_1(s)$ used to be called as semi-Poissonian
statistics.\\

 Roughly,  particles in the free-flow regime
do not interact between each other, contrary to the synchronized
states when the interaction leads to the particle repulsion and to
the fact that $\lim_{s \rightarrow 0^+} P_\nu(s)=0$ for positive
$\nu.$ One of aims of this article is to answer whether it is
possible to get a synchronized state for the asymmetric simple
exclusion model by an appropriate choice of input parameters, or
the model is in the free regime for all parameters.\\

\section{ASEP with open boundaries}

Consider sites located on a chain of length $N$. Each site $i$
($1\leq i\leq N$) is either occupied by a particle, or
empty. During the time interval $dt$, each particle jumps to the
next site (in the defined direction) with probability $dt$ if the
target site is empty and does not hop otherwise. Moreover, during
the time interval $dt$, a particle can enter the first chain
position $(i=1)$ with probability $\alpha\,dt$ $(\alpha \in
\langle 0,1\rangle)$ if this position is empty and a particle at
the site $i=N$ can leave the chain with probability $\beta\,dt$
$(\beta \in \langle 0,1\rangle)$ if this site is occupied. These
rules define a simple traffic model with a hard-core repulsion
which  decreases the probability for
occurrence of the two particles close each other. This was
\textit{de facto} the reason why we have chosen ASEP for our
investigation.\\
Two alternative methods of exact solution of the model are outlined in
\cite{Derrida1}. We will use the \textit{matrix method} in our
computations. Consider the matrices $D,E$ and vectors $\langle w
|$ and $| v \rangle$ satisfying the following algebraic rules
\begin{equation}DE=D+E \label{Rule1}\end{equation}
\begin{equation} \langle w |E=\frac{1}{\alpha}\langle w
|\label{Rule2}\end{equation}
\begin{equation}D| v \rangle=\frac{1}{\beta}| v
\rangle\label{Rule3}\end{equation} Then we can determine the
probability of the arbitrary configuration
$\mathcal{C}=(n_1,m_1,n_2,m_2,\ldots)$ of the model through the
formula \begin{equation}P(\mathcal{C})=\frac{1}{Z_N} \,
\frac{\langle w |D^{n_1}E^{m_1}D^{n_2}E^{m_2}\ldots | v
\rangle}{\langle w | v \rangle}\label{Config},\end{equation} where
$n_1$ is the number of the empty sites, starting with the entering
site of the chain (and $(n_1+1)$-th site is occupied), $m_1$ is
the number of occupied sites starting with the site number $n_1+1$
(and $(n_1+m_1+1)$-th site is empty), and so on. $Z_N$ is a normalization
constant. Hence $\sum_i n_i + \sum_j m_j =N.$ Owing to the fact

\begin{equation}(D+E)^n=\sum_{m=1}^{n}
\frac{m\,(2n-m-1)!}{n!(n-m)!} \sum_{i=0}^{m} E^i D^{m-i}
\label{D+E}\end{equation}

 we can rewrite the normalization factor $Z_N$ to the form

 \begin{equation}Z_N\equiv\frac{\langle w|
(D+E)^N| v \rangle}{\langle w| v \rangle} = \sum_{m=1}^N
\frac{m\,(2N-m-1)!}{N!(N-m)!} \sum_{i=0}^m
\left(\frac{1}{\alpha}\right)^i \,
\left(\frac{1}{\beta}\right)^{m-i} \label{ZN}\end{equation}

Hence,

\begin{equation}Z_N=\sum_{m=1}^N \frac{m\,(2N-m-1)!}{N!(N-m)!}
\frac{\beta^{-m-1} - \alpha^{-n-1}}{\beta^{-1} - \alpha^{-1}}
\label{ZN2}\end{equation}

These are the known results  \cite{Derrida1} which will be used
below.

\section{Headway distribution in the ASEP (a special case)}

Our aim is an analytical expression for
the spacing distribution, which means
the probability of the gap of the length $k$ behind the particle
being on the $i$-th site. One can determine this probability
$P^{(i)}(k)$ from the equation (\ref{Config}) as follows
\begin{equation}
P^{(i)}(k)=\frac{\mathcal{A}}{Z_N\,\langle w| v \rangle} \,
\langle w| (D+E)^{i-1} D\,E^{k-1} D\,(D+E)^{N-i-k}\,|v\rangle,
\label{gap}
\end{equation}
since the first $i-1$ sites could be arbitrary occupied or not,
$i$-th site is occupied, the following $k-1$ sites are empty, the
site $i+k$ is occupied and the rest sites could be
arbitrary occupied or not. If these rules are satisfied, one can
see the gap of the length $k$ behind the $i$-th position.
The constant $\mathcal{A}$ can be calculated from the condition
$$\sum_{k=1}^{N-i} P^{(i)}(k)=1 
$$

Generally, the matrices $D,E$ are infinite-dimensional, but
provided the condition \begin{equation} \alpha + \beta =1
\label{special-case} \end{equation}  one can find
$D,E$ to be real numbers. The equations (\ref{Rule2}) and (\ref{Rule3})
imply that $D=1/\beta,E=1/\alpha$ and the rule (\ref{Rule1})
imposes that ${}^1/{}_\alpha + {}^1/{}_\beta = {}^1/{}_{\alpha
\beta}$, which leads to the equation (\ref{special-case}).
Instituting (\ref{special-case}) and (\ref{ZN}) into the
(\ref{gap}) we obtain
\begin{equation}
P^{(i)}(k)=\mathcal{A}\,\alpha^2\,\beta^{k-1} \, (k \geq 1)
\label{headway-special} \end{equation}

Strictly speaking, the derivation of the formula
(\ref{headway-special}) from the equation (\ref{gap}) is possible
only for $k\geq3$ since for $k=1,2$ the relation (\ref{gap}) has a
different form. But it can easily be verified that
(\ref{headway-special}) holds true also for $k=1,2$. Note now that
the result (\ref{headway-special}) is independent of the value
$i$. For large $N$, after the re-scaling to $\langle s \rangle=1,$
equation (\ref{headway-special}) changes to the form $P(s):=
P(k-1)=e^{-s}$, which is the Poissonian distribution. This clearly
shows that under the condition (\ref{special-case}), the
interaction among the particles vanishes in the continuous limit
 and hence the movement of
the particles is asymptotically free (in the sense mentioned in the section
$2$).\\ Also the real headway distribution (without the
re-scaling) can be determined in this section. If
(\ref{special-case}) holds true then
\begin{equation} P^{(i)}(k)=\mathcal{N}
\left(\frac{1}{\alpha}
\right)^{N-2}\,\left(\frac{1}{\beta}\right)^{N+1-k},
\label{real-gap}
\end{equation} when the normalization pre-factor $\mathcal{N}$ can be
computed either from the condition $\sum_{k=1}^{N-i} P^{(i)}(k)
=1$ (through all the lengths of the gap) or from the equation
$$\frac{1}{\mathcal{N}}=\frac{\langle
w|(D+E)^{N-i}D\,(D+E)^{N-i}|v\rangle\,-\,\langle
w|(D+E)^{i-1}D\,E^{N-i}|v\rangle}{\langle w|v \rangle},$$ which
takes into account all the possible combinations with occupied
site number $i$ and rejects the combinations when all the sites
(behind the site number $i$) are empty. This provides

$$\frac{1}{\mathcal{N}}=\left(\frac{1}{\alpha}\right)^{N-i}
\left(\frac{1}{\beta}\right)^{N}(1-\beta^{N-i}),$$

which together with (\ref{real-gap}) yields $$
P^{(i)}(k)=\frac{\alpha}{1-\beta^{N-i}}\,\beta^{k-1}
$$

In the large $N$ limit, one can calculate from
(\ref{special-case}) that $$ P^{(i)}(k)=\alpha\,\beta^{k-1}
$$ and hence $$ \langle k \rangle \equiv \sum_{k=1}^\infty k\,
P^{(i)}(k)=\alpha\,\sum_{k=1}^\infty
k\,\beta^{k-1}=\frac{1}{\alpha} $$\label{mean-gap}

for the mean headway.

\section{Headway distribution of the ASEP (general case)}

In this section we are solving the ASEP (equation (\ref{gap}),
strictly speaking) for arbitrary values of the parameters
$\alpha,\, \beta.$ Introduce now the following pre-factors

$$\omega=\omega(m,k,i,N,p)=\frac{m\,p\,(2N-2i-2k-m-1)!(2i-p-3)!}{(N-i-k)!(N-i-k-m)!(i-1)!(i-p-1)!}$$

$$ A_1=A_1(p,q,a,w)=\left( \begin{array}{c} p-q-a+w-1
\\ p-q-1 \end{array} \right)
$$

$$ A_2=A_2(p,q,b,w)=\left( \begin{array}{c} p-q-b+w-1
\\ w-1 \end{array} \right)
$$

With the help of them and the computations presented in an
Appendix we obtain the desired result

\inspicture

\begin{equation}
P^{(i)}(k)=\frac{\mathcal{A}}{Z_N} \sum_{p=1}^{i-1}
\sum_{m=1}^{N-i-k} \sum_{l=0}^m \,\omega\, (S_1+S_2+S_3+S_4+S_5)
\label{Finalk}
\end{equation}

where

$$S_1=\left(\frac{1}{\beta}+l \right)\sum_{q=0}^{p-1} \left(
\frac{1}{\alpha} \right)^q\left( \frac{1}{\beta}
\right)^{p-q+m-l+1}$$

$$S_2=\sum_{q=0}^{p-1} \sum_{w=1}^{k-1}\left(\, \sum_{a=1}^w
A_1\left( \frac{1}{\alpha} \right)^{q+a}\left( \frac{1}{\beta}
\right)^{m-l+1} + \sum_{b=1}^{p-q} A_2\left( \frac{1}{\alpha}
\right)^q\left( \frac{1}{\beta} \right)^{m+b-l+1} \right)$$

$$S_3=l\,\sum_{q=0}^{p-1} \sum_{w=1}^{z+k-1}\left(\, \sum_{a=1}^w
A_1\left( \frac{1}{\alpha} \right)^{q+a}\left( \frac{1}{\beta}
\right)^{m-l} + \sum_{b=1}^{p-q} A_2\left( \frac{1}{\alpha}
\right)^q\left( \frac{1}{\beta} \right)^{m+b-l} \right)$$

$$S_4=\left( \frac{1}{\alpha} \right)^p\left( \frac{1}{\beta}
\right)^{m-l+2}+\sum_{w=1}^{k-1} \left( \frac{1}{\alpha}
\right)^{p+w}\left( \frac{1}{\beta} \right)^{m-l+1}$$

$$S_5=l\left( \frac{1}{\alpha} \right)^p\left( \frac{1}{\beta}
\right)^{m-l+1}+l\sum_{w=1}^{z+k-1} \left( \frac{1}{\alpha}
\right)^{p+w}\left( \frac{1}{\beta} \right)^{m-l}$$

The equation (\ref{Finalk}) presents the final result of this
article. In the special case (\ref{special-case}) the equations
(\ref{Finalk}) and (\ref{real-gap}) provide the same values, which
supports the correctness of our result.  Now, one can show that
the headway distribution of the ASEP (after re-scaling) is of a
Poissonian type for any parameters $\alpha,\beta$ (see Figure 1).

\section{Conclusion}

To conclude, the ASEP shows the Poissonian behavior for all
possible values of the parameters $\alpha$ and $\beta$. Thus, this
model can simulate only the traffic in free-flow regime, which
means that
\begin{equation} \lim_{s \rightarrow 0^+} P(s) \gg 0
\label{ConditionP} \end{equation} To be precise, we have to
express that as the parameter $\beta$ goes to zero ($\beta
\rightarrow 0^+$) the Poissonian distribution changes since one
can easily to follow that $P(s)=\delta(s)=\delta(k-1)$ for
$\beta=0$ and $\alpha\neq0$. This holds true due to the shift of
the probability $P(s):=P(k-1)$. The above mentioned shift is done
because $P(k)=0$ for every $k<1$ in the original model. However,
for every $\beta$ we get the headway distribution satisfying the
condition (\ref{ConditionP}).\\

On contrary, in \cite{Gier} a different variant of the ASE-model
(high speed traffic model with open boundaries) is presented. The
particles of this model also enter the system at the left and
leave at the right of the chain. Moreover, all particles will move
with their maximal possible speed, which is given by the speed
limit $v_{max}$ (a maximal length of the jump). For $v_{max}=1$
one can obtain the model investigated in this article. Numerical
simulations of the high speed model with $v_{max}\geq2$ clearly
show that the gap distribution can start from the origin (i.e.
$\lim_{s \rightarrow 0^+} P(s) = 0$) for some values
$\alpha,\beta.$\\ Hence, the high speed model has the synchronized
states as well, on contrary to the asymmetric simple exclusion
process and it can be used to the full simulation of the real-road
traffic.

\section{Acknowledgement}
Author is very grateful to V.B. Priezzhev and Petr \v Seba for
valuable advice and for the help with writing of this paper.

\section{Appendix}

At the beginning, we cite some useful lemmas.

\vspace{0.5 cm} \underline{Lemma 1}

 \vspace{0.5 cm} Let $m,n,a$ are the natural numbers and $a \leq m.$ Then
\begin{equation} \sum_{i=a}^m \left( \begin{array}{c} n-i+m-1 \\
n-1 \end{array} \right) = \left( \begin{array}{c} n+m-a \\ n
\end{array} \right) \label{lemma1} \end{equation}

\textit{Proof:} by means of the mathematical induction for the
numbers $m$ and $n$ \\

\vspace{0.5 cm} \underline{Lemma 2}

 \vspace{0.5 cm} Assume that $D,E$ are arbitrary matrices
 fulfilling the equation $DE=D+E.$ Let $m,n$ are the natural
 numbers. Then

\begin{equation} D^nE^m = \sum_{i=1}^m \left( \begin{array}{c} n-i+m-1 \\
n-1 \end{array} \right) \, E^i + \sum_{j=1}^n  \left(
\begin{array}{c} n-j+m-1 \\ m-1
\end{array} \right) \, D^j \label{lemma2} \end{equation}

\textit{Proof:} by means of the mathematical induction for the
numbers $m$ and $n$ and (\ref{lemma1})

\vspace{0.5 cm}


Now we can rewrite the equation (\ref{gap}) in order to use the
rules (\ref{Rule2}),(\ref{Rule3}). Thus, it is necessary to
replace the matrices $D$ on the right-hand side and matrices $E$
to the left-hand side of the equation (\ref{gap}). The expression
(\ref{D+E}) yields

$$(D+E)^{N-i-k} = \sum_{m=1}^{N-i-k}
\frac{m\,(2N-2i-2k-m-1)!}{(N-i-k)!(N-i-k-m)!} \sum_{l=0}^m
E^lD^{m-l} 
$$

Thus,

 \begin{eqnarray} D\,(D+E)^{N-i-k} = \sum_{m=1}^{N-i-k}
\frac{m\,(2N-2i-2k-m-1)!}{(N-i-k)!(N-i-k-m)!} \sum_{l=0}^m
D\,E^lD^{m-l}=  \nonumber \\ = \left| \begin{array}{c} D\,E^l = D
+ \sum_{z=1}^l E^z
\\ D\,E^lD^{m-l} = D^{m-l+1} + \sum_{z=1}^l E^zD^{m-l} \end{array}
\right| =  \nonumber \\ = \sum_{m=1}^{N-i-k}
\frac{m\,(2N-2i-2k-m-1)!}{(N-i-k)!(N-i-k-m)!} \sum_{l=0}^m \left(
D^{m-l+1} + \sum_{z=1}^l E^zD^{m-l} \right)  \nonumber 
\end{eqnarray}

It can be verified that \begin{eqnarray} E^{k-1}D\,(D+E)^{N-i-k}=
\nonumber \\ =\,\sum_{m=1}^{N-i-k}
\frac{m\,(2N-2i-2k-m-1)!}{(N-i-k)!(N-i-k-m)!} \sum_{l=0}^m
\left(E^{k-1}D^{m-l+1} + \sum_{z=1}^l E^{z+k-1}D^{m-l}\right)
\nonumber 
\end{eqnarray}

Furthermore,

 \begin{eqnarray} D\,E^{k-1}D\,(D+E)^{N-i-k}=
 \left|\begin{array}{c} D\,E^{k-1} = D + \sum_{w=1}^{k-1} E^w
\\ D\,  E^{z+k-1} = D + \sum_{w=1}^{z+k-1} E^w \end{array}
\right| =  \nonumber \\ = \sum_{m=1}^{N-i-k}
\frac{m\,(2N-2i-2k-m-1)!}{(N-i-k)!(N-i-k-m)!} \nonumber \\
\sum_{l=0}^m \left( D^{m-l+2} + \sum_{w=1}^{k-1}  E^wD^{m-l+1} +
l\,D^{m-l+1} + \,l\, \sum_{w=1}^{z+k-1} E^wD^{m-l}\right)
\label{vlozeno}
\end{eqnarray}

Another use of the expression (\ref{D+E}) gives \begin{equation}
(D+E)^{i-1}= \sum_{p=1}^{i-1} \frac{p\,(2i-p-3)!}{(i-1)!(i-p-1)!}
\sum_{q=0}^p E^qD^{p-q} \label{112} \end{equation}

Let now \begin{equation}
\gamma=\gamma(m,k,i,N)=\frac{m\,(2N-2i-2k-m-1)!}{(N-i-k)!(N-i-k-m)!}
\label{113a} \end{equation}

\begin{equation}
\delta=\delta(i,p)=\frac{p\,(2i-p-3)!}{(i-1)!(i-p-1)!}
\label{113b}
\end{equation}

Applying (\ref{113a}),(\ref{113b}), you can transform the
equations (\ref{vlozeno}) and (\ref{112}) to

\begin{eqnarray} D\,E^{k-1}D\,(D+E)^{N-i-k}=\nonumber \\ =\sum_{m=1}^{N-i-k}
 \sum_{l=0}^m \gamma \left( D^{m-l+2} + \sum_{w=1}^{k-1}  E^wD^{m-l+1} +
 l\,
D^{m-l+1} + \,l\, \sum_{w=1}^{z+k-1} E^wD^{m-l}\right)
\nonumber 
\end{eqnarray}

$$ (D+E)^{i-1}=\sum_{p=1}^{i-1} \sum_{q=0}^p \delta \,E^qD^{p-q}
$$

Hence, \begin{eqnarray} (D+E)^{i-1}D\,E^{k-1}D\,(D+E)^{N-i-k} =
\sum_{p=1}^{i-1} \sum_{q=0}^p \sum_{m=1}^{N-i-k} \sum_{l=0}^m
\delta \, \gamma \nonumber \\ \left( E^qD^{p-q+m-l+2} +
\sum_{w=1}^{k-1} E^q \overbrace{D^{p-q}E^w} D^{m-l+1} \, +
\nonumber \right. \\ \left. l\, E^qD^{p-q+m-l+1} \, + \,l\,
\sum_{w=1}^{z+k-1} E^q \overbrace{D^{p-q}E^w} D^{m-l} \right)
\label{115}
\end{eqnarray}

In the expression (\ref{115}) we can change the products indicated
by the horizontal braces. Using (\ref{lemma2}) one can obtain

\begin{equation} D^{p-q}E^w =\sum_{a=1}^w \left(
\begin{array}{c} p-q-a+w-1 \\ p-q-1 \end{array} \right) E^a +
\sum_{b=1}^{p-q} \left(
\begin{array}{c} p-q-b+w-1 \\ w-1 \end{array} \right) D^b
\label{116} \end{equation}

Note that (\ref{116}) is valid only for $p>q.$ The remaining
arrangement is the substitution of the (\ref{116}) into the
equation (\ref{115}). With the help of the pre-factors

$$ A_1=\left( \begin{array}{c} p-q-a+w-1
\\ p-q-1 \end{array} \right) 
$$

$$ A_2=\left( \begin{array}{c} p-q-b+w-1
\\ w-1 \end{array} \right) 
$$

the expression (\ref{115}) turns into the form

\begin{eqnarray}
(D+E)^{i-1}D\,E^{k-1}D\,(D+E)^{N-i-k} = \sum_{p=1}^{i-1}
\sum_{q=0}^{p-1} \sum_{m=1}^{N-i-k} \sum_{l=0}^m \delta\,\gamma
\nonumber \\ \left(  E^qD^{p-q+m-l+2} + \,l\, E^qD^{p-q+m-l+1} +
\sum_{w=1}^{k-1} \sum_{a=1}^w A_1 E^{q+a} D^{m-l+1}\right. \, +
\nonumber
\\ + \left. \sum_{w=1}^{k-1} \sum_{b=1}^{p-q} A_2 E^q D^{m+b-l+1} \,+\,
l\, \sum_{w=1}^{z+k-1} \sum_{a=1}^w A_1 E^{q+a}D^{m-l} \,+\,l\,
\sum_{w=1}^{z+k-1} \sum_{b=1}^{p-q} A_2 E^qD^{b+m-l} \right) +
\nonumber \\ + \sum_{p=1}^{i-1} \sum_{m=1}^{N-i-k} \sum_{l=0}^m
\delta\,\gamma \left( E^pD^{m-l+2}+\sum_{w=1}^{k-1}
E^{p+w}D^{m-l+1} + l\,E^pD^{m-l+1} + \,l\, \sum_{w=1}^{z+k+1}
E^{p+w}D^{m-l} \right) \nonumber 
\end{eqnarray}

which together with the rules (\ref{Rule2}),(\ref{Rule3}) yields

\begin{center}
\begin{eqnarray}
P^{(i)}(k) =\frac{\mathcal{A}}{Z_N} \sum_{p=1}^{i-1}
\sum_{q=0}^{p-1} \sum_{m=1}^{N-i-k} \sum_{l=0}^m  \delta \, \gamma
\left\{ \left( \frac{1}{\alpha} \right)^q\left( \frac{1}{\beta}
\right)^{p-q+m-l+2} + \,l\, \left( \frac{1}{\alpha}
\right)^q\left( \frac{1}{\beta} \right)^{p-q+m-l+1} \right. \,+
\nonumber
\\ + \sum_{w=1}^{k-1} \sum_{a=1}^w A_1 \left( \frac{1}{\alpha}
\right)^{q+a} \left( \frac{1}{\beta} \right)^{m-l+1} \,  +
\sum_{w=1}^{k-1} \sum_{b=1}^{p-q} A_2 \left( \frac{1}{\alpha}
 \right)^q \left( \frac{1}{\beta} \right)^{m+b-l+1} \,+ \nonumber
 \\ + \left.
\,l\, \sum_{w=1}^{z+k-1} \sum_{a=1}^w A_1\left( \frac{1}{\alpha}
\right)^{q+a}\left( \frac{1}{\beta} \right)^{m-l} \,+\,l\,
\sum_{w=1}^{z+k-1} \sum_{b=1}^{p-q} A_2\left( \frac{1}{\alpha}
\right)^q\left( \frac{1}{\beta} \right)^{b+m-l} \right\} +
\nonumber \\ + \frac{\mathcal{A}}{Z_N} \sum_{p=1}^{i-1}
\sum_{m=1}^{N-i-k} \sum_{l=0}^m \delta\,\gamma \left\{ \left(
\frac{1}{\alpha} \right)^p\left( \frac{1}{\beta}
\right)^{m-l+2}+\sum_{w=1}^{k-1} \left( \frac{1}{\alpha}
\right)^{p+w}\left( \frac{1}{\beta} \right)^{m-l+1} \,\right.+
\nonumber
\\ \left.+ \,l\,\left( \frac{1}{\alpha} \right)^p\left( \frac{1}{\beta}
\right)^{m-l+1} + \,l\, \sum_{w=1}^{z+k+1} \left( \frac{1}{\alpha}
\right)^{p+w}\left( \frac{1}{\beta} \right)^{m-l} \right\}
\end{eqnarray}
\end{center}

\end{document}